\documentclass[aps,twocolumn,superscriptaddress,prl]{revtex4-1}

\usepackage{graphicx}
\usepackage{color}
\usepackage{amssymb}
\usepackage{amsmath}
\usepackage{subfigure}
\usepackage{siunitx}
\usepackage{hyperref}
\usepackage{accents}

\newcommand{\PRLsection}[1]{\noindent {\it#1} -}
\newcommand{\op}[1]{\hat{#1}}
\newcommand{\agg}[1]{\hat{#1}^{\dag}}
\newcommand{\media}[1]{\langle #1 \rangle}
\newcommand{\E}[2]{\op{a}_{ #1} (#2)}
\newcommand{\Ea}[2]{\agg{a}_{ #1} (#2)}
\newcommand{\R}[3]{R_{#1,#2}^{ (2)} (#3)}
\newcommand{\Rs}[2]{R_{#1,#2}^{ (2)} }

\newcommand{\myvec}[1]{{\bf #1}}

\newcommand{\nC}{{\Phi_C}}
\newcommand{\nS}{{\Phi_S}}

\usepackage{color}
 \definecolor{BLACK}{gray}{0}
 \definecolor{WHITE}{gray}{1}
 \definecolor{RED}{rgb}{1,0,0}
 \definecolor{GREEN}{rgb}{0,.75,0}
 \definecolor{BLUE}{rgb}{0,0,1}
 \definecolor{CYAN}{cmyk}{1,0,0,0}
 \definecolor{MAGENTA}{cmyk}{0,1,0,0}
 \definecolor{YELLOW}{cmyk}{0,0,1,0}

\makeatletter

\begin{document}

\title{
Optical spin squeezing: bright beams as high-flux entangled photon sources }

\author{Federica A. Beduini}
\email[Corresponding author: ]{federica.beduini@icfo.es}
\affiliation{ICFO-Institut de Ciencies Fotoniques, Mediterranean Technology Park, 08860 Castelldefels (Barcelona), Spain}

\author{Morgan W. Mitchell}
\affiliation{ICFO-Institut de Ciencies Fotoniques, Mediterranean Technology Park, 08860 Castelldefels (Barcelona), Spain}
\affiliation{ICREA-Instituci\'{o} Catalana de Recerca i Estudis Avan\c{c}ats, 08015 Barcelona, Spain}

\date{\today}

\begin{abstract}

In analogy with the spin-squeezing inequality of Wang and Sanders [Physical Review A {\bf 68}, 012101 (2003)], we find inequalities describing macroscopic polarization correlations that are obeyed by all classical fields, and whose violation implies entanglement of the photons that make up the optical beam. We consider a realistic and exactly-solvable experimental scenario employing polarization-squeezed light from an optical parametric oscillator (OPO) and find {polarization entanglement for {postselected} photon pairs separated by less than the OPO coherence time}.  The entanglement is robust against losses and extremely bright: efficiency can exceed that of existing ``ultra-bright'' pair sources by at least an order of magnitude.   {This translation of spin-squeezing inequalities to the optical domain will enable direct tests of discrete variable entanglement in a squeezed state.} 
\end{abstract}

\maketitle

\PRLsection{Introduction}
{
Spin squeezing inequalities, in which squeezing of an ensemble implies entanglement of the constituent particles, are powerful tools for understanding the relationship between macroscopic and microscopic quantum features \cite{GuhnePR2009}. The work of S{\o}rensen {et al.} \cite{SorensenN2001} showed that squeezing of a spin-1/2 ensemble implies   entanglement in the ensemble.  Wang and Sanders \cite{WangPRA2003}  considered symmetric spin systems  and showed that squeezing implies entanglement in every reduced two-atom density matrix.  Similar results have been found for larger-spin systems, for other kinds of squeezing, and for multi-partite entanglement \cite{VitaglianoPRL2011,Korbicz2005,KorbiczPRA2006,GuhnePR2009}.  Spin squeezing has been produced in a number of experiments \cite{AppelPNAS2009,LerouxPRL2010,GrossN2010,ChenPRL2011,SewellPRL2012}, implying entanglement of macroscopic numbers of atoms.  To date, there has been no direct observation of the implied  entanglement.  

Here we present a result analogous to that of Wang and Sanders, but for optical fields.  To our knowledge, this is the first spin-squeezing-type inequality in the optical domain {i.e., the first demonstration that optical continuous-variable (CV) non-classicality implies discrete variable (DV) entanglement} { upon projection to photon pairs}.  Production and detection of optical squeezing is a well-developed technology, with quadrature squeezing levels reaching 12.3 dB \cite{MehmetOE2011}.  Simultaneously, efficient detection of photons is routine in quantum optics laboratories, as is quantum state tomography of entangled pairs \cite{JamesPRA2001,AdamsonPRL2007}.  Together, these offer the possibility to test the predicted relations between macroscopic squeezing and microscopic entanglement.  

We also give a practical implementation and show that the efficiency of narrowband entangled pair generation exceeds the state-of-the-art by at least an order of magnitude, a promising result for quantum networking.

~

\newcommand{\DM}{{\cal R}}
\newcommand{\mymatrix}[1]{\accentset{\leftrightarrow}{#1}}
\newcommand{\mytensor}[1]{\bar{\bar{#1}}}
\newcommand{\SymOp}{\pi_V}
\PRLsection{Scenario}
\label{sec:condition}
We consider a beam with a single-spatial mode and stationary statistical properties, i.e., a continuous-wave beam.   We suppose that the $H$ and $V$ modes of this beam are in a product state, with $H$ being a strong coherent state, and $V$ being a weak non-classical state.  We also assume the state is invariant under $\SymOp \equiv \exp[i \pi a^\dagger_V a_V]$, i.e., a $\pi$ phase shift of the $V$ mode or equivalently $a_V \rightarrow - a_V$.  This includes an important class of practical nonclassical states, for example squeezed thermal states and even cat states.  Because of the product structure, there is no entanglement of the $H$, $V$ modes.  The coherent state contains many photons but no entanglement, while the entanglement content of the non-classical state is limited, due to its low brightness.  

~

\PRLsection{Nonclassicality and entanglement}  The nonclassicality and entanglement properties are related through the first- and second-order correlation functions
\begin{equation}
 R^{(1)}_{i,j}(\tau) \equiv \media{\Ea{i}{t}\,\E{j}{t+\tau}}   
 \label{eq:G1}
\end{equation}
and
\begin{equation}
  \R{ij}{mn}{\tau}\equiv\media{\Ea{i}{t}\Ea{j}{t+\tau}\E{n}{t+\tau}\E{m}{t}},
   \label{eq:R}
\end{equation}
respectively,  where $\op{a}_{i}$ is the annihilation operator for the mode $i \in \{H,V\}$.  Note that we have inverted the last two indices in the definition of $R^{(2)}$, in keeping with the convention in photonic quantum state tomography \cite{JamesPRA2001}.

A simple non-classicality condition is found using the Cauchy-Schwarz inequality \footnote{For example, with a pure classical state $\R{HH}{VV}{\tau} = \left<\alpha^*_H(t)\alpha^*_H(t+\tau)\alpha_V(t+\tau)\alpha_V(t) \right>$ where $\alpha_{H,V}$ are c-numbers.  Rearranging we find $\R{HH}{VV}{\tau} = \left< \alpha^*_H(t)\alpha_V(t+\tau) \alpha^*_H(t+\tau)\alpha_V(t) \right>$ which is constrained by inequality (\ref{eq:G2HHVV}) after similar RHS rearrangements. } {or by systematic derivations \cite{ShchukinPRL2005,VogelPRL2008}}:  classical fields  obey 
\begin{subequations}
	\begin{eqnarray}
		|\R{HH}{VV}{\tau}|^2 &\le& \R{HV}{HV}{\tau}\R{VH}{VH}{\tau}
		\label{eq:G2HHVV}  \\
		|\R{HV}{VH}{\tau}|^2 &\le& \R{HH}{HH}{\tau} \R{VV}{VV}{\tau}
		\label{eq:G2HVHV}
	\end{eqnarray}
\end{subequations}
whereas quantum fields can violate these inequalities.  {We now show that Eqs. (\ref{eq:G2HHVV}),(\ref{eq:G2HVHV}) imply polarization entanglement when DV detection methods, i.e. photon counting, are used.}

\newcommand{\Rate}{W}
For single-photon detection with polarization $\myvec{p}$, a vector in the $H,V$ basis, Glauber theory indicates the average rate of detections as
\begin{equation}
	\Rate^{(1)}_\myvec{p} = {\rm Tr}[\Pi_\myvec{p} \, \myvec{R}^{(1)}(0)]\,,
\end{equation}
where $\Pi_\myvec{p} = \myvec{p} \wedge \myvec{p}$ is a projector onto $\myvec{p}$. 
It is conventional to define the one-photon {\it observable density matrix} (ODM) $\myvec{\mathcal{R}}^{(1)} \equiv \myvec{R}^{(1)}/{\rm Tr}[\myvec{R}^{(1)}]$, so that relative probabilities of detection are given by the Born rule  
\begin{equation}
	P^{(1)}_\myvec{p} = {\rm Tr}[\Pi_\myvec{p} \, \myvec{\mathcal{{R}}}^{(1)}(0)]\,.
\end{equation}

In the same way, the pair polarization $\myvec{r} \equiv \myvec{p}\otimes \myvec{q}$, 
where $\myvec{p}$, $\myvec{q}$ are unit polarization vectors, gives the rate at which photon pairs arrive separated by time $\tau$
\begin{equation}
	\Rate^{(2)}_\myvec{r}(\tau) d\tau = {\rm Tr}[\Pi_\myvec{r} \, \myvec{R}^{(2)}(\tau)] d\tau \,.
\end{equation}
Again, this is usually expressed via the Born rule $P^{(2)}_\myvec{r}(\tau) = {\rm Tr}[\Pi_\myvec{r} \, \myvec{\mathcal{R}}^{(2)}{(\tau)}]\,,$ where $\myvec{\mathcal{R}}^{(2)}(\tau) \equiv \myvec{R}^{(2)}(\tau)/{\rm Tr}[\myvec{R}^{(2)}(\tau)]$ is the two-photon ODM.  Recovery of $\myvec{\mathcal{R}}^{(N)}$ from $P^{(N)}$ is the subject of quantum state tomography \cite{JamesPRA2001}. 
}

\newcommand{\SymOp}{\pi_V}

Using the $\SymOp$ invariance, we find an ODM similar to that found for symmetric atomic states \cite{WangPRA2003}:
\begin{equation}
   \mathcal{R}^{ (2)} \propto
   \begin{pmatrix}
      \Rs{HH}{HH} &   0   &   0   &   \Rs{HH}{VV} \\
      0 &   \Rs{HV}{HV} &   \Rs{HV}{VH}  &   0\\
      0 &    \Rs{VH}{HV} &   \Rs{VH}{VH}  &   0\\
      \Rs{VV}{HH}   &   0   &   0   &   \Rs{VV}{VV} \\
   \end{pmatrix}\;, 
   \label{eq:matrix}
\end{equation}
where all elements are functions of $\tau$.  This describes a mixture of a state in the 
\{$HH,VV$\} subspace and another in \{$HV,VH$\}.

By inspection, positivity under partial transposition of the above density matrix is equivalent to the two conditions of eq.~\eqref{eq:G2HHVV},\eqref{eq:G2HVHV}.  Thus non-classical polarization correlations imply {DV} entanglement of the photons in the extracted modes, for stationary beams with the $\pi_V$ symmetry described above. 
{In the following section, we will describe a feasible experimental scenario that allows to violate the spin-squeezing-type inequalities with available technologies.} 

~

\PRLsection{Practical implementation} 
Continuous wave non-classical polarizations have been produced by combining two bright squeezed beams of orthogonal polarization \cite{KorolkovaPRA2002,BowenPRL2002}, by optical self-rotation \cite{RiesPRA2003} and by combining a coherent state ($H$-polarized) with $V$-polarized squeezed vacuum \cite{Predojevic2008}.  We consider the last case, which manifestly shows symmetry under $\SymOp$.

For the squeezed vacuum, we consider a sub-threshold OPO, as described by Collett and Gardiner \cite{Collett1984}.  
The field
operator $\op{a}_V$ is expressed via a Bogoliubov transformation of the vacuum input and loss 
reservoir operators $\op{a}_1$ and $\op{a}_2$, respectively   
\begin{equation}
\label{Bogoliubov}
		\begin{split}
     \E{V}{\omega} = & f_1 (\omega)\,\E{1}{\omega} + f_2 (\omega)\,\Ea{1}{-\omega} \\ 
      &+ f_3 (\omega)\,\E{2}{\omega} + f_4 (\omega)\,\Ea{2}{-\omega}.
     \end{split}
\end{equation}
The coefficients
\begin{subequations}
\begin{eqnarray}
   f_1(\omega) &=&\left[\eta^2-(1-\eta-i\omega/\delta\nu)^2+\mu^2\right]A^{-1}(\omega)\;,\\
   f_2(\omega) &=&\left[2\,\eta\,\mu\right]A^{-1}(\omega)\;,\\
   f_3(\omega) &=&\left[2\sqrt{\eta\,(1-\eta)}(1-i\omega/\delta\nu)\right]A^{-1}(\omega)\;,\\
   f_4(\omega) &=&\left[2\mu\sqrt{\eta\,(1-\eta)}\right]A^{-1}(\omega)\;,\\
   A(\omega) &=&(1-i\omega/\delta\nu)^2-\mu^2\;,\\
   \end{eqnarray}
\end{subequations}
are functions of the experimental parameters of the OPO: the cavity FWHM
bandwidth $\delta\nu$, the photon flux of the squeezed vacuum
state $\nS= R^{(1)}_{\text{V,V}}(0) = \frac{\mu^2 \eta \delta\nu}{1 - \mu^2}$,  and the cavity escape coefficient $\eta$, i.e. the ratio between the transmission of the output coupler $T_1$ and the sum of both the intracavity losses $T_2$ and the transmission of the output coupler ($\eta=T_1(T_1+T_2)^{-1}$). $\mu^2$ is the pump power expressed as a fraction of the threshold power.  We take the phase of the pump field equal to $0$ for simplicity.

\newcommand{\var}{{\rm var}}

The time-domain correlation functions required for ~\eqref{eq:R} can
be computed as Fourier integrals, to find 
\begin{subequations}
\begin{eqnarray}
    \Rs{HH}{HH} &=& \nC^2 \\
    \Rs{HV}{HV} &=&  \nC \,  \alpha\mu\,,\\
    \Rs{HH}{VV} (\tau)&=& \nC \, \alpha  \left[\cosh x + \mu \sinh x \right]e^{-\delta\nu|\tau|} \\
   \label{eq:elements}
  \Rs{HV}{VH} (\tau)&=&  \nC \, \alpha\left[\mu\cosh x+\sinh x\right] e^{-\delta\nu|\tau|}  \\
    \Rs{VV}{VV} (\tau) &=& \alpha^2 \left\{  \beta +\left[(1-\mu^2)\cosh(2x) \right. \right.  \nonumber \\
    & & \left. \left. +2\mu\sinh(2x)\right] e^{-2\delta\nu|\tau|}\right\} 
\end{eqnarray}
\end{subequations}
{where}
\begin{subequations}
\begin{eqnarray}
 x &=& \mu \delta\nu |\tau|\;, \\
   \alpha &=&\frac{\eta\mu\delta\nu }{1-\mu^2}\;,\\
   \beta &=& \frac{1}{\pi\delta\nu(1-\mu^2)}\left\{\mu^4(1-\eta-\pi\delta\nu) +\right.\nonumber\\
    & &\left.+ \mu^2[\pi\delta\nu+2\eta(1+\eta)-1]  +6\eta^2-9\eta+4\right\}
\end{eqnarray}
\end{subequations}
{and $\nC= R^{(1)}_{\text{H,H}}(0)$ is the photon flux of the
coherent state. }

~

\PRLsection{Entanglement under realistic conditions}
\label{sec:numerical} We now show that it is possible to achieve
either high entanglement or high rates of entangled pairs with feasible experimental values.

We quantify the entanglement associated with a pair extracted from a polarization
squeezed state by means of the concurrence~\cite{Wootters1998}
\begin{equation}
   \mathcal{C} = \max(0,\sqrt{\lambda_1} -\sqrt{\lambda_2} -\sqrt{\lambda_3} -\sqrt{\lambda_4})\;,
\end{equation}
where $\lambda_i$ are the eigenvalues of $\mathcal{R}^{(2)}(\tau)[\sigma_y
\otimes \sigma_y][\mathcal{R}^{(2)}]^*[\sigma_y \otimes \sigma_y]$  in
decreasing order and $\sigma_y$ is a Pauli matrix. 
The relevant experimental parameters are the time interval between detections $\tau$ and the average photon fluxes of the coherent and squeezed state, $\nC$ and $\nS$ respectively. Changing the other parameters do not change significatively the concurrence, so we fix them to typical experimental values, specifically  $\delta\nu = 8$~MHz and $\eta = 0.93$, from  \cite{Predojevic2008}.

\begin{figure}[t]
   \includegraphics[width=.9\linewidth]{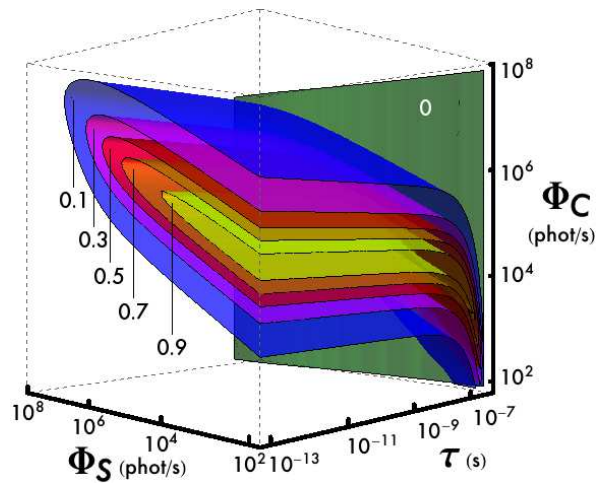}
   \caption{
  (color online) Entanglement of photon pairs within a polarization-squeezed state.  State contains squeezed vacuum from a sub-threshold OPO and an orthogonally-polarized coherent state as described in text.  Contours show concurrence ${\cal C}$ versus average photon fluxes in the coherent ($\Phi_C$) and squeezed ($\Phi_S$) beams,  and versus time separation $\tau$.    Fixed experimental parameters: cavity linewidth $\delta\nu = 8$~MHz and cavity escape coefficient $\eta=0.93$.  
  }
   \label{fig:concurrence}
\end{figure}

 Figure~\ref{fig:concurrence} shows that the state is entangled for any choice of $\nC$ and $\nS$, provided that the two photons are detected within the coherence time of the squeezed state, while the concurrence goes to zero when $\tau \gtrsim 1/\delta\nu$. However, the concurrence does not change much in a wide range of time separation $\tau$, which goes from the time resolution of actual single photon detectors (some tens of picoseconds) to hundreds of nanoseconds ($\approx 1/\delta \nu$). 

\begin{figure}[t]
   \includegraphics[width=1.0\linewidth]{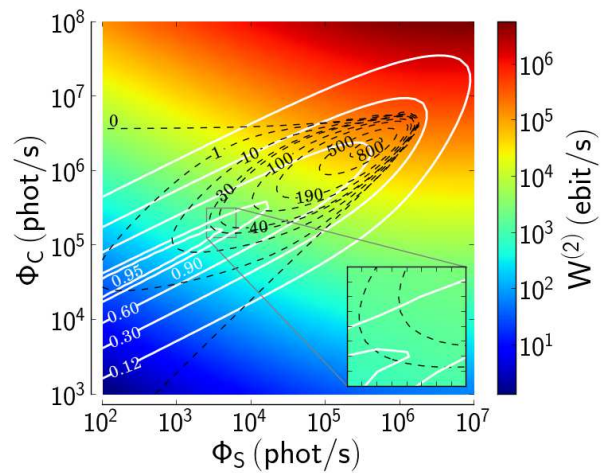}
      \caption{(color online) Total concurrence flux $W^{(2)}$ versus input fluxes $\Phi_C$ and $\Phi_S$. {Solid white contours show concurrence $\mathcal{C}$ for $\tau$=1 ns.  Dashed black contours show non-locality figure of merit $\beta$ (see text) for $\Delta\tau$=1 ns. Inset: a region where entanglement dilution increases $\beta$.}}
   \label{fig:ebits}
\end{figure}

We estimate the entangled pair flux by averaging the concurrence with the corresponding photon flux:
\begin{equation}
	W^{(2)} = \int_{-\infty}^{+\infty} d\tau \mathrm{Tr}[\myvec{R}^{(2)}(\tau)] \, \mathcal{C}(\tau),
\end{equation}
and we plot it in Fig.~\ref{fig:ebits} compared to concurrence: the experimental parameters $\nC$ and $\nS$ can be suitably chosen in order to obtain a Bell-like state with high concurrence ($\mathcal{C}\geq 0.9$ inside the innermost (yellow)  surface in Fig.~\ref{fig:concurrence}).

However, there are some cases where high entanglement flux can be more important than maximal entanglement.  For example, non-maximally entangled spin-1/2 states which violate a Bell inequality can be useful for teleportation~\cite{HorodeckiPhysLettA1996}. A ``typical'' state statisfying this requirements
\begin{equation}
   \mathcal{R}^{ (2)}_{\rm typ} \approx
   \begin{pmatrix}
      0.601 &   0   &   0   &  0.388 \\
      0 &  0.067 &  0.067 &   0\\
      0 &  0.067 &  0.067 &   0\\
       0.388   &   0   &   0   &   0.264 \\
   \end{pmatrix}\;, 
   \label{eq:TypicalTPDM}
\end{equation}
obtained with squeezed beam flux $\Phi_S=2\cdot10^5$~photons/s (2.6\% OPO threshold), coherent beam flux $\Phi_C=2\cdot10^6$~photons/s and arrival-time difference $\tau = 1$~ns, 
   can combine a high rate of entangled pairs  with easily detectable concurrence: the state of Eq. (\ref{eq:TypicalTPDM}) has $\mathcal{C}=0.64$ and $W^{(2)} = 7\cdot10^5$ ebit/s, well above the $3\cdot10^4$ ebits/s that can be reached by states with high concurrence ($\mathcal{C}\geq0.9)$. 
    Such a state can {be used for teleportation with up to 88$\%$ fidelity~\cite{Hu2012} and can} be generated feasibly with current technology: in fact, it only needs 2.3~dB of squeezing, well within existing capabilities.
    
{The ability to trade brightness against entanglement purity may be advantageous also in applications of quantum non-locality.  Hu et al.~\cite{Hu2012} calculate the achievable Clauser-Horne-Shimony-Holt inequality violation $\Delta S\equiv S -2$ for states with the form of $\mathcal{R}^{ (2)}$.  Using that result, and the fact that statistical significance (in standard deviations) scales as $(T \Phi_{\Delta\tau})^{1/2}$, where $T$ is the acquisition time and $\Phi_{\Delta\tau} \approx  \mathrm{Tr}[\myvec{R}^{(2)}(0)] \Delta \tau$ is the rate of detections within a coincidence window of width $\Delta \tau \ll \delta \nu^{-1}$, we find the figure of merit $\beta \equiv \Delta S^2 \Phi_{\Delta\tau}$ to describe how quickly a Bell inequality violation acquires statistical significance.  As shown in Fig.~\ref{fig:ebits}, the largest $\beta$ occur for bright, modestly-entangled states, ${\cal C} \approx 0.6$, and in some regions  entanglement dilution (increasing $\Phi_C$ while keeping $\Phi_S$ constant) increases $\beta$.  
}


~
\newcommand{\ratio}{R_{P/S}}

\PRLsection{Comparison to other sources} 
DV polarization entanglement is the preferred embodiment for many quantum information tasks, e.g. free-space quantum key distribution \cite{SteinlechnerOE2012} and optical quantum computing \cite{WaltherN2005}.   { Our source incorporates a coherent state, which gives a high brightness but also less strict photon-number correlations than true pair sources such as parametric down-conversion.   A conservative estimate of the pairs-to-singles flux ratio is $\ratio \equiv W^{(2)}/(\Phi_S + \Phi_C)$.  As an example, $\Phi_S = 2 \cdot 10^5$, $\Phi_C = 2 \cdot 10^6$, as in Eq. (\ref{eq:TypicalTPDM}), gives $\ratio = 0.32$ pairs per photon, whereas an ideal entangled pair source would have $\ratio =1/2$.  This imperfect correlation would prevent a loophole-free Bell test with this state, for example.  In contrast, the brightness is attractive for} quantum networking with atoms, an application currently
limited by the spectral brightness of narrow-band sources 
\cite{HaaseOL2009,PiroNP2011}.  Cavity-enhanced sources of polarization-entangled photons have demonstrated detected (inferred) spectral brightness {per pump power} of 70 (1221) pairs/(s MHz mW)  \cite {WolfgrammOE2008,WolfgrammJOSAB2010} and 50 (5500) pairs/(s MHz mW)  \cite{ZhangNP2011}.  The spectral brightness we predict here is $\sim 1.2 \cdot 10^5$/(s MHz) for an 8 MHz bandwidth.  The required pump power for the OPO (about 10 \% of threshold) depends on the implementation: for a doubly-resonant OPO, e.g. \cite{BowenPRL2002}, the power is $\sim$ \SI{5}{mW}, while for a triply-resonant OPO, e.g.  \cite{MartinelliJOA2001}, it can be $\sim$\SI{50}{\micro W}. The source described here is thus one to three orders of magnitude brighter than existing sources.  

~

\PRLsection{Discussion}
Even though the polarization squeezed state is a product of an entangled state (squeezed vacuum) and a classical one (coherent) with orthogonal polarization, our result shows that the contribution of both initial states is fundamental for the pairwise entanglement of the final state. In fact, the maximum concurrence corresponds to the case that most resembles a Bell state, in which it is equally probable to detect two $H$-polarized or two $V$-polarized photons ($\mathcal{R}_{HH,HH}^{(2)}(\tau)\approx\mathcal{R}_{VV,VV}^{(2)}(\tau)\approx 0.5)$, showing that the coherent state plays an important role in the generation of polarization entangled pairs.

~

\PRLsection{Conclusions} 
We have derived a ``spin-squeezing inequality'' for photons, analogous to the result of Wang and Sanders \cite{WangPRA2003} for spins in symmetric states.  The result shows that non-classical macroscopic polarization correlations imply microscopic entanglement of the photons in the beam.  Considering polarization-squeezed light from a sub-threshold OPO, we find exact expressions for the entangled state, and show that an experimental demonstration of {DV} entanglement {associated to squeezing}, not yet practical with atoms, is feasible with photons.  We predict {polarization-}entangled photon sources robust against losses and brighter than existing ``ultra-bright'' sources by one to three orders of magnitude, of considerable interest for quantum networking applications.

\begin{acknowledgments}
We thank Jaroslaw Korbicz, Alejandra Valencia, and Philipp Hauke for helpful discussions. This work was supported by the Spanish MINECO under the project MAGO (Ref. FIS2011-23520), by the European Research Council  under the project AQUMET {and by Fundaci\'{o} Privada CELLEX}. \end{acknowledgments}

\bibliographystyle{apsrev4-1}
\bibliography{BigBib130123}

\end{document}